**Optimal allocation of attentional resource to multiple items with unequal relevance**

Nuwan de Silva and Wei Ji Ma

Center for Neural Science and Department of Psychology, New York University

**In natural perception, different items (objects) in a scene are rarely equally relevant to the observer. The brain improves performance by directing attention to the most relevant items, for example the ones most likely to be probed. For a general set of probing probabilities, it is not known how attentional resources should be allocated to maximize performance. Here, we investigate the optimal strategy for allocating a fixed resource budget $E$ among $N$ items when on each trial, only one item gets probed. We develop an efficient algorithm that, for any concave utility function, reduces the $N$-dimensional problem to a set of $N$ one-dimensional problems that the brain could plausibly solve. We find that the intuitive strategy of allocating resource in proportion to the probing probabilities is in general not optimal. In particular, in some tasks, if resource is low, the optimal strategy involves allocating zero resource to items with a nonzero probability of being probed. Our work opens the door to normatively guided studies of attentional allocation.**

**INTRODUCTION**

Objects in natural scenes rarely have the same relevance. Given limited resources, it is crucial to better encode objects that are more relevant. In the laboratory, the process of allocating resources can be investigated by cueing one item to which attention should be directed, for example by starting a trial by pointing a small arrow at that item's location (e.g. (Dosher & Lu, 2000; Eckstein, Shimozaki, & Abbey, 2002; Eriksen & James, 1986; Folk, Remington, & Johnston, 1992; Gobell & Carrasco, 2005; Kinchla, Chen, & Evert, 1995; Posner, 1980; Prinzmetal, Ha, & Khani, 2010; Theeuwes, 1991)). Then, the array of $N$ items is presented. Afterwards, one item is "probed", i.e. the subject is asked to report a property of that item. The cued item is probed with a higher probability than non-cued items; this probability is called "cue validity" and is typically approximately known to the subject. It is consistently found that cueing improves the performance at the cued location. However, the performance benefits of cueing are typically not quantitatively related to the value of cue validity. One exception was the recent work by Paul Bays (Bays, 2014), who in the context of a cued delayed-estimation task, examined how the gains of neuronal subpopulations *should* be set in order to maximize performance, when the total gain is held constant.

In attentional cueing experiments, the probing probabilities of the non-cued item are typically equal to each other. For example, when $N$=4, the vector of probing probabilities could be (0.6, 0.1, 0.1, 0.1). This is, however, a rather restricted class of probing probability vectors.

Here we prove several results on how to optimally allocate a fixed resource budget among multiple items, for a large class of tasks and general probing probabilities.

**TASKS**
We assume that the observer strives to maximize some form of expected utility. The class of tasks that we consider is defined by following three properties:

1. **Separability:** the utility function can be written as a sum of terms each of which only depends on the amount of resource allocated to a particular item.
2. **Equivalence:** Each of these local utility functions has, up to a multiplicative constant, the same functional dependence on the amount of resource allocated.
3. **Diminishing returns:** the local utility function is concave in the amount of resource.

The first two conditions are satisfied by many cuing paradigms in which the stimuli are generated independently and the subject is interrogated about one of them, including commonly used cued discrimination (Rahnev et al., 2011) and cued estimation (Bays, 2014; Zhang & Luck, 2008) paradigms. The last condition is satisfied by most utility functions in psychology and neuroscience, whether experimentally imposed or assumed in models.
 For the class of tasks defined by the three conditions, and when resource is a non-negative real-valued variable, we will prove that given any set of probing probabilities, there exists a unique optimal allocation. Then, we will derive an algorithm that is guaranteed to find this optimal allocation in an amount of computational time that is linear in the number of stimuli. It turns out that in general, the intuitive strategy of allocating attention in proportion to the probing probabilities ("proportional allocation") is not optimal. In particular, when the derivative of the local utility function at zero is finite, and the amount of resource is low enough, optimal allocation involves "dropping" – i.e., allocating zero resource to items that have a nonzero probability of being probed. On the other hand, if this derivative is infinite, optimal allocation does not involve dropping.

**PROBLEM STATEMENT**
We consider tasks of the following nature. The experimenter presents $N$ independently generated items that are all characterized by a scalar variable, for example orientation. We denote the values of this variable by $s_1, s_2, \ldots, s_N$. The brain makes noisy measurements of these values, denoted by $x_1, x_2, \ldots, x_N$, respectively. A resource parameter $\kappa_i$ controls the noise level of the $i^{th}$ measurement: the lower $\kappa_i$, the noisier the corresponding measurement (the notation $\kappa$ comes from the common notation for the concentration parameter of a Von Mises distribution). The experimenter probes one of the items, say the $L^{th}$ one, and the subject is tasked with estimating a deterministic or random variable $T$ that depends on $s_L$ but not on the other $s_i$'s. For example, in an orientation estimation task, $T$ would be $s_L$ itself, while in an orientation categorization task, $T$ would be the category from which $s_L$ is drawn. The subject uses the measurement $x_L$ to make an estimate $\hat{T}$ of $T$. The distribution of $\hat{T}$ given $T$ and $\kappa_i$, which we denote by $p(\hat{T}|T;\kappa_i)$, therefore depends on the noise model.

The probability that the experimenter probes the $i^{th}$ item is $p_i$. Without loss of generality, we assume that $p_1 \geq p_2 \geq \ldots \geq p_N$. We also assume that the observer possesses perfect knowledge of all $p_i$.

We further assume that the observer possesses a measure $A(T,\hat{T})$ of the utility of the estimate on a given trial. For example, in an orientation estimation task, we would have $T=s_L$ and $\hat{T}=\hat{s}_L$; then, $A(T,\hat{T})$ could be the negative magnitude of the angle between $T$ and $\hat{T}$. In general, the expected "local" utility across trials when the $i^{th}$ item gets probed is

$$u(\kappa_i) = \sum_{T,\hat{T}} A(T,\hat{T}) p(T,\hat{T};\kappa_i) \tag{1}$$

Overall expected utility across all trials is therefore

$$U(\kappa) = \sum_{i=1}^{N} p_i u(\kappa_i) \tag{2}$$

where $\kappa=(\kappa_1, \ldots, \kappa_N)$. For brevity, we will from now on refer to expected local utility and overall expected utility as "local utility" and "overall utility", respectively. Because the expected value in Eq. (1) is taken with respect to the distribution $p(\hat{T}|T;\kappa_i)$, which depends on the noise model, the utility function also depends on the noise model.

Now we are ready to state the problem. Suppose the brain has control over the resource vector $\kappa$ subject to the constraint that the resource budget (total amount of resource) equals $E$:

$$\sum_{i=1}^{N} \kappa_i = E$$

Then, we can think of – and we will indeed refer to – each possible $\kappa$ as an attention allocation strategy, or allocation for short. We wish to find the allocation $\kappa$ that maximizes utility $U(\kappa)$.

**INTUITION BEHIND THE SOLUTION**
At first sight, one might think that it is optimal to allocate attention in proportion to the items' probing probabilities. For example, when $N=2$, if one item has 70% probability of being probed and the other item 30%, then the proportional allocation strategy would allocate $0.7E$ and $0.3E$, respectively to each item. It turns out that this strategy is not necessarily optimal. Instead, the optimal strategy depends on the derivative of the utility function. To get an intuition for why this is the case, we show in Figure 1 an example of a utility function. Assume that the first item has a probability $p_1$, and the second item a probability $p_2=1-p_1$ of being probed. One way to allocate a resource budget $E$ is to allocate all of it to the first item. From Eq. (2), the overall utility of this allocation is $U(E,0)=p_1 u(E)+p_2 u(0)$. To determine whether this allocation is locally optimal, we ask what happens if we were to take a small amount of resource $\delta$ away from the first item and allocate it to the second item. Then, $U$ would change to $U(E-\delta,\delta) = p_1 u(E-\delta) + p_2 u(\delta)$. In the limit when $\delta$ is small, the change in $U$ is $-p_1 \delta u'(E) + p_2 \delta u'(0)$. Therefore, if $p_1 u'(E) < p_2 u'(0)$, allocating all resource to the first item is not optimal: the gain obtained from investing some resource in the second item outweighs the loss in the first item. However, if $p_1 u'(E) > p_2 u'(0)$, then allocating all resource to the first item is locally optimal (and as we will show below, also globally optimal) and it is optimal to completely "drop" the second item from consideration. This example makes clear that the probing probabilities and the derivative of the utility function together determine the optimal allocation.

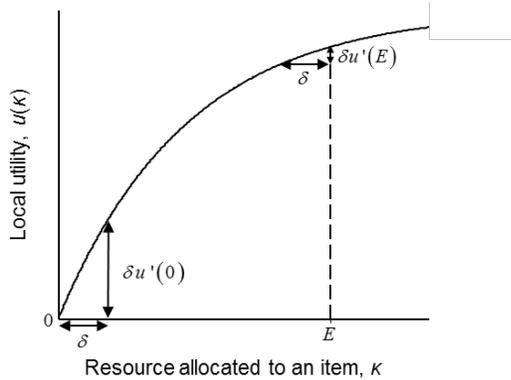

**Figure 1.** Intuition behind the condition $\lambda^{(k-1)} < p_{k+1} f(0)$ in the algorithm of Theorem 1 to find the optimal allocation when N=2.

**SOLUTION**

We now consider the common case that resource $\kappa$ is a non-negative, real-valued variable, such as inverse variance. Moreover, we assume that $u$ is a concave and increasing function of $\kappa$. This reflects diminishing returns, which is a feature of many utility functions: incrementing $\kappa$ by a fixed amount increases utility by less as $\kappa$ is bigger. For example, this is true in a fine discrimination task, where $u$ is proportion correct and $\kappa$ is the inverse variance of the noise corrupting the stimuli that have to be discriminated. The following theorem gives an algorithm to find the globally best strategy to allocate attention.

**Theorem 1:** Assume that $\sum_{i=1}^{N} \kappa_i = E$, and that $u$ is concave and increasing on $[0,E]$. Denote the derivative of $u$ by $f$. Then, the global optimum of $U(\kappa) = \sum_{i=1}^{N} p_i u(\kappa_i)$ is

$$\kappa^{(\text{opt})} = \left( \kappa_1^{(\text{opt})}, ..., \kappa_{k_{\max}}^{(\text{opt})}, 0, ..., 0 \right), \qquad (3)$$

where
$$\kappa_i^{(\text{opt})} = f^{-1}\left( \frac{\lambda^{(k_{\max}-1)}}{p_i} \right) \qquad (4)$$

and $k_{\max}$ and $\lambda^{(k_{\max}-1)}$ are given by the following algorithm.

$$\begin{aligned}
&\lambda^{(0)} \equiv p_1 f(E); \\
&k = 1; \\
&\text{while } \lambda^{(k-1)} < p_{k+1} f(0) \text{ and } k < N \\
&\quad \lambda^{(k)} = \text{solve}\left( g^{(k)}(\lambda) = E \right); \\
&\quad k = k+1; \\
&\text{end} \\
&k_{\max} = k; \\
&\text{return } k_{\max}, \lambda^{(k_{\max}-1)}
\end{aligned} \qquad (5)$$

The algorithm starts with $k=1$ and checks whether or not $p_1 f(E) < p_2 f(0)$ holds; this corresponds to the intuition in the previous section. If the answer is no, then the optimal allocation vector is $(E, 0, ..., 0)$ and the algorithm terminates. If the answer is yes, then the algorithm solves $g^{(1)}(\lambda) = E$ to find $\lambda^{(1)}$, sets $k$ to 2, and checks whether or not $\lambda^{(1)} < p_3 f(0)$. Proceeding in this way, the algorithm increases $k$ by 1 until the condition $\lambda^{(k-1)} < p_{k+1} f(0)$ does not hold. As soon as this happens, the algorithm terminates, and the current value of $k$ becomes the number of non-zero elements in the optimal allocation vector. This algorithm is not only efficient, it also illuminates the nature of the optimal solution, as captured by the condition $\lambda^{(k-1)} < p_{k+1} f(0)$. In particular, it indicates the circumstances under which dropping items from attention is optimal.

**PROOF OF THEOREM 1**
To prove Theorem 1, we will first reduce the problem to an optimization in a lower-dimensional space. Let $u$ be a concave function. For $k \in \{1, ..., N-1\}$, we define the function

$$U^{(k)}(\kappa_1,...,\kappa_k) \equiv p_{k+1}u(E-\kappa_1-...-\kappa_k) + \sum_{i=1}^{k} p_i u(\kappa_i)$$
$$= U(\kappa_1,...,\kappa_k, E-\kappa_1-...-\kappa_k, 0,...,0) \quad (6)$$

on the $k$-dimensional closed region

$$V^{(k)} \equiv \left\{ (\kappa_1,...,\kappa_k) \in [0,E]^k \,\middle|\, \sum_{i=1}^{k} \kappa_i \leq E \right\} \quad (7)$$

Thus, $U^{(k)}(\kappa)$ is the overall utility of an allocation $\kappa$ in which only the $k+1$ items most likely to be probed receive resource, and all other items receive zero resource. We next define the $(k-1)$-simplex

$$S^{(k)} \equiv \left\{ (\kappa_1,...,\kappa_k) \in [0,E]^k \,\middle|\, \sum_{i=1}^{k} \kappa_i = E \right\} \quad (8)$$

Then $S^{(k)}$ is a subset of the boundary of $V^{(k)}$.

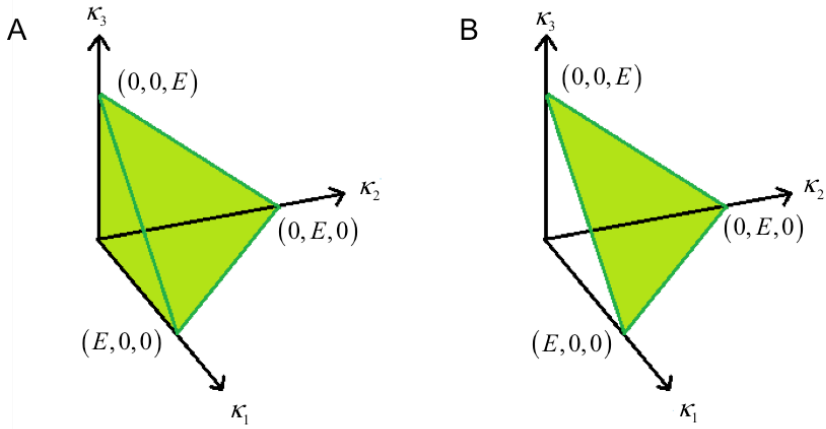

**Figure 2.** Illustration of the region $V^{(k)}$ and the simplex $S^{(k)}$ given by Eqs. (7) and (8), for $k=3$. **(A)** The region $V^{(3)}$. **(B)** The 2-simplex $S^{(3)}$.

We note that
$$\bar{I}(\kappa_1,...,\kappa_{N-1}) = (\kappa_1,...,\kappa_{N-1}, E-\kappa_1-...-\kappa_{N-1})$$

is a one-to-one and onto mapping from $V^{(N-1)}$ to $S$. Also

$$U^{(N-1)}(\kappa_1,...,\kappa_{N-1}) = p_N u(E - \kappa_1 - ... - \kappa_{N-1}) + \sum_{i=1}^{N-1} u(\kappa_i) p_i$$
$$= U(\kappa_1,...,\kappa_{N-1}, E - \kappa_1 - ... - \kappa_{N-1})$$
$$= U(\overline{I}(\kappa_1,...,\kappa_{N-1}))$$

Therefore, the optimum $\kappa^{(\text{opt})}$ of $U$ in $S$ relates to the optimum $\kappa^{(N-1)}$ of $U^{(N-1)}$ in $V^{(N-1)}$ by $\kappa^{(\text{opt})} = \overline{I}(\kappa^{(N-1)})$.

Since the components of each element of $S$ add up to $E$, finding the optimum of $U$ in $S$ is equivalent to finding the optimum of $U^{(N-1)}(\kappa_1,...,\kappa_{N-1}) = U(\kappa_1,...,\kappa_{N-1}, E - \kappa_1 - ... - \kappa_{N-1})$ in $V^{(N-1)}$:

$$\begin{aligned} \kappa_i^{(\text{opt})} &= \kappa_i^{(N-1)} \text{ for } i \in \{1,...,N-1\} \\ \kappa_N^{(\text{opt})} &= E - \kappa_{N-1}^{(N-1)} - ... - \kappa_1^{(N-1)} \end{aligned} \quad (9)$$

Now we will prove three lemmas about the quantities we just defined.

**Lemma 1:** $U^{(k)}$ is concave in $V^{(k)}$ for $k \in \{1,...,N-1\}$.

**Proof of Lemma 1:** Clearly, $V^{(k)}$ is a convex set. Let $\kappa^{(1)} = (\kappa_1^{(1)},...,\kappa_k^{(1)}) \in V^{(k)}$ and $\kappa^{(2)} = (\kappa_1^{(2)},...,\kappa_k^{(2)}) \in V^{(k)}$. Since $V^{(k)}$ is convex, we have $\kappa^{(1)} + \lambda(\kappa^{(2)} - \kappa^{(1)}) \in V^{(k)}$ for any $\lambda \in [0,1]$, and

$$U^{(k)}(\kappa^{(1)} + \lambda(\kappa^{(2)} - \kappa^{(1)})) = p_{k+1} u\left(E - \sum_{i=1}^{k}(1-\lambda)\kappa_i^{(1)} + \lambda\kappa_i^{(2)}\right) + \sum_{i=1}^{k} p_i u((1-\lambda)\kappa_i^{(1)} + \lambda\kappa_i^{(2)})$$
$$= p_{k+1} u\left((1-\lambda)\left(E - \sum_{i=1}^{k}\kappa_i^{(1)}\right) + \lambda\left(E - \sum_{i=1}^{k}\kappa_i^{(2)}\right)\right)$$
$$+ \sum_{i=1}^{k} p_i u((1-\lambda)\kappa_i^{(1)} + \lambda\kappa_i^{(2)})$$

Since the local utility function $u$ is concave,

$$U^{(k)}\left(\boldsymbol{\kappa}^{(1)} + \lambda\left(\boldsymbol{\kappa}^{(2)} - \boldsymbol{\kappa}^{(1)}\right)\right) \geq p_{k+1}(1-\lambda)u\left(E - \sum_{i=1}^{k}\kappa_i^{(1)}\right) + p_{k+1}\lambda u\left(E - \sum_{i=1}^{k}\kappa_i^{(2)}\right)$$

$$+ \sum_{i=1}^{k} p_i(1-\lambda)u\left(\kappa_i^{(1)}\right) + \lambda p_i u\left(\kappa_i^{(2)}\right)$$

and

$$U^{(k)}\left(\boldsymbol{\kappa}^{(1)} + \lambda\left(\boldsymbol{\kappa}^{(2)} - \boldsymbol{\kappa}^{(1)}\right)\right) \geq (1-\lambda)\left[p_{k+1}u\left(E - \sum_{i=1}^{k}\kappa_i^{(1)}\right) + \sum_{i=1}^{k} p_i u\left(\kappa_i^{(1)}\right)\right]$$

$$+ \lambda\left[p_{k+1}u\left(E - \sum_{i=1}^{k}\kappa_i^{(2)}\right) + \sum_{i=1}^{k} p_i u\left(\kappa_i^{(2)}\right)\right]$$

$$U^{(k)}\left(\boldsymbol{\kappa}^{(1)} + \lambda\left(\boldsymbol{\kappa}^{(2)} - \boldsymbol{\kappa}^{(1)}\right)\right) \geq (1-\lambda)U^{(k)}\left(\boldsymbol{\kappa}^{(1)}\right) + \lambda U^{(k)}\left(\boldsymbol{\kappa}^{(2)}\right) \qquad (10)$$

Therefore, $U^{(k)}$ is concave in $V^{(k)}$. □

We are interested in the optimal resource allocation, i.e. the vectors $\boldsymbol{\kappa}$ that maximize utility. Thus, we need to find the critical points of $U^{(k)}$ in $V^{(k)}$.

**Lemma 2:** $\kappa^*$ is a critical point of $U^{(k)}$ in $V^{(k)}$ if and only if

$$\kappa_i^* = f^{-1}\left(\frac{\lambda^{(k)}}{p_i}\right) \text{ for } i \in \{1,2,...,k\} \qquad (11)$$

where $f = u'$, and $\lambda^{(k)}$ is the solution to

$$g^{(k)}(\lambda) = E, \qquad (12)$$

where

$$g^{(k)}(\lambda) \equiv \sum_{i=1}^{k+1} f^{-1}\left(\frac{\lambda}{p_i}\right) \text{ for } k \in \{1,...,N-1\} \qquad (13)$$

**Proof of Lemma 2**: The partial derivative of $U^{(k)}$ with respect to $\kappa_i$ is

$$\frac{\partial}{\partial \kappa_i} U^{(k)}(\kappa_1,...,\kappa_k) = p_{k+1}\frac{\partial}{\partial \kappa_i} u(E - \kappa_1 - ... - \kappa_k) + p_i \frac{\partial}{\partial \kappa_i} u(\kappa_i)$$

$$= p_i u'(\kappa_i) - p_{k+1} u'(E - \kappa_1 - ... - \kappa_k)$$

If $\kappa^*$ is a critical point, then

$$p_i u'(\kappa_i^*) = p_{k+1} u'(E - \kappa_1^* - ... - \kappa_k^*) \equiv \lambda^{(k)} \text{ for } i \in \{1,...,k\} \quad (14)$$

Since $u$ is concave and increasing, $u' = f$ is invertible in $[0, E]$ and Eq. (14) gives

$$\kappa_i^* = f^{-1}\left(\frac{\lambda^{(k)}}{p_i}\right) \text{ for } i \in \{1,...,k\} \quad (15)$$

and

$$E - \kappa_1^* - ... - \kappa_k^* = f^{-1}\left(\frac{\lambda^{(k)}}{p_{k+1}}\right) \quad (16)$$

Combining Eq. (15) for $i \in \{1,...,k\}$ with Eq. (16) gives the implicit equation $g^{(k)}(\lambda) = E$.
To show the converse, assume $g^{(k)}(\lambda^{(k)}) = E$ and the elements of $\kappa^*$ defined by Eq. (15). From Eq. (15), it follows that

$$E - \sum_{i=1}^{k}\kappa_i^* = E - \sum_{i=1}^{k} f^{-1}\left(\frac{\lambda^{(k)}}{p_i}\right).$$

Since $E = g^{(k)}(\lambda^{(k)})$, we can use Eq. (13) to find that the right-hand side is equal to $f^{-1}\left(\frac{\lambda^{(k)}}{p_{k+1}}\right)$.

It follows that

$$f\left(E - \sum_{i=1}^{k}\kappa_i^*\right) = \frac{\lambda^{(k)}}{p_{k+1}}$$

and $p_i u'(\kappa_i^*) = p_{k+1} u'(E - \kappa_1^* - ... - \kappa_k^*) = \lambda^{(k)}$ for $i \in \{1,...,k\}$. Therefore, all partial derivatives will equal zero at $\kappa^*$. □

**Lemma 3:** If the optimum $\kappa^{(k)}$ of $U^{(k)}$ in $V^{(k)}$ belongs to the boundary of $V^{(k)}$, then it belongs to the subset

$$S^{(k)} = \left\{ (\kappa_1,...,\kappa_k) \in [0,E]^k \,\Big|\, \sum_{i=1}^{k} \kappa_i = E \right\} \quad (17)$$

of the boundary of $V^{(k)}$.

**Proof of Lemma 3**: Let $\kappa^{(k)}$ belong to the boundary of $V^{(k)}$. Suppose $\kappa^{(k)}$ does not belong to $S^{(k)}$. Then $\sum_{i=1}^{k} \kappa_i^{(k)} < E$ and $\kappa_j^{(k)} = 0$ for some $j \in \{1,...,k\}$. Define an alternative allocation $\kappa^{(\text{alt})}$ by

$$\kappa_i^{(\text{alt})} = \kappa_i^{(k)} \text{ for } i \neq j$$
$$\kappa_j^{(\text{alt})} = E - \sum_{i \neq j} \kappa_i^{(k)} > 0$$

Then $\kappa^{(\text{alt})} \in S^{(k)}$ belongs to the boundary of $V^{(k)}$ and

$$U^{(k)}\left(\kappa^{(\text{alt})}\right) - U^{(k)}\left(\kappa^{(k)}\right) = p_{k+1}\left[u\left(E - \kappa_1^{(\text{alt})} - ... - \kappa_k^{(\text{alt})}\right) - u\left(E - \kappa_1^{(k)} - ... - \kappa_k^{(k)}\right)\right] + p_j\left[u\left(\kappa_j^{(\text{alt})}\right) - u\left(\kappa_j^{(k)}\right)\right]$$
$$= p_{k+1}\left[u(0) - u\left(\kappa_j^{(\text{alt})}\right)\right] + p_j\left[u\left(\kappa_j^{(\text{alt})}\right) - u(0)\right]$$
$$> 0$$

So $\kappa^{(k)}$ cannot be the optimum of $U^{(k)}$ in the boundary of $V^{(k)}$. □

We will need the following well-known theorem in the derivations below.

**Theorem 2** (Roberts & Varberg, 1973)**:** Let $f$ be a concave function in the convex set $S$. Let $x$ be an interior point of $S$. The $x$ is the global optimum of $f$ in $S$ if and only if $x$ is a critical point of $f$.

Intuitively, this states that for convex and concave functions, local optima are global optima. Combining Lemmas 1 through 3 and Theorem 2, we obtain the following result.

**Proposition 1:** Denote the global optimum of $U^{(k)}$ in $V^{(k)}$ by $\kappa^{(k)}$. If $g^{(k)}(\lambda)=E$ has a solution $\lambda^{(k)}$, then $\kappa^{(k)}=\kappa^*$, with the components of $\kappa^*$ given by Eq. (11), is the global optimum of $U^{(k)}$ in $V^{(k)}$. Otherwise,

$$\kappa^{(k)} = \left(\kappa_1^{(k-1)},...,\kappa_{k-1}^{(k-1)}, E - \kappa_1^{(k-1)} - ... - \kappa_{k-1}^{(k-1)}\right) \quad (18)$$

where

$$\boldsymbol{\kappa}^{(k-1)} = \left( \kappa_1^{(k-1)}, \ldots, \kappa_{k-1}^{(k-1)} \right)$$

is the global optimum of $U^{(k-1)}$ in $V^{(k-1)}$.

**Proof of Proposition 1**: If $g^{(k)}(\lambda) = E$ has a solution $\lambda^{(k)}$, then from Lemma 2, $\kappa^*$ with components given by Eq. (11) is a critical point of $U^{(k)}$ in $V^{(k)}$. Since $V^{(k)}$ is a convex set and $U^{(k)}$ is concave in $V^{(k)}$, it follows from Theorem 2 that $\kappa^*$ is the global optimum of $U^{(k)}$ in $V^{(k)}$.

On the other hand, if $g^{(k)}(\lambda) = E$ does not have a solution, then, from Lemma 1, $U^{(k)}$ does not have a critical point in $V^{(k)}$. Then, it follows from Theorem 2 that the global optimum of $U^{(k)}$ in $V^{(k)}$ cannot be in the interior of $V^{(k)}$. Therefore, the global optimum must be in the boundary of $V^{(k)}$. Then, from Lemma 3, the global optimum must be in the subset $S^{(k)}$ of the boundary of $V^{(k)}$. Let $(\kappa_1, \ldots, \kappa_k) \in S^{(k)}$. Then

$$\begin{aligned} U^{(k)}(\kappa_1, \ldots, \kappa_k) &= p_{k+1} u\left( E - \kappa_1 - \ldots - \kappa_{k-1} - (E - \kappa_1 - \ldots - \kappa_{k-1}) \right) \\ &\quad + \sum_{i=1}^{k-1} u(\kappa_i) p_i + p_k u(E - \kappa_1 - \ldots - \kappa_{k-1}) \\ &= p_{k+1} u(0) + U^{(k-1)}(\kappa_1, \ldots, \kappa_{k-1}) \end{aligned}$$

Therefore, the optimum $\kappa^{(k)}$ of $U^{(k)}$ in $S^{(k)}$ relates to the global optimum $\kappa^{(k-1)}$ of $U^{(k-1)}$ in

$$V^{(k-1)} = \left\{ (\kappa_1, \ldots, \kappa_{k-1}) \in [0, E]^{k-1} \mid \sum_{i=1}^{k-1} \kappa_i \leq E \right\}$$

through Eq. (18). □

Proposition 1 directly leads to the following theorem for finding the maximum of $U$ in $S$.

**Theorem 3:** Let

$$m = \max\left\{ k \in \{1, \ldots, N-1\} \mid g^{(k)}(\lambda) = E \text{ has a solution } \lambda^{(k)} \right\} \quad (19)$$

and let $\kappa^{(\text{opt})}$ be the global optimum of $U$ in $S$. Then,

(i) If $1 \leq m \leq N-1$, then the components of $\kappa^{(\text{opt})}$ are given by

$$\kappa_i^{(opt)} = \begin{cases} f^{-1}\left(\dfrac{\lambda^{(m)}}{p_i}\right) & \text{for } i \in \{1,...,m+1\} \\ 0 & \text{for } i \in \{m+2,...,N\} \end{cases} \quad (20)$$

(ii) If $g^{(k)}(\lambda)=E$ does not have a solution for any $k \in \{1,...,N-1\}$, then the global optimum is

$$\boldsymbol{\kappa}^{(opt)} = (E,0,...,0), \quad (21)$$

**Proof of Theorem 3:**
(i) $1 \leq m \leq N-1$: Since $g^{(m)}(\lambda)=E$ has the solution $\lambda^{(m)}$, it follows that the components of $\kappa^{(m)}$ are

$$\kappa_i^{(m)} = f^{-1}\left(\dfrac{\lambda^{(m)}}{p_i}\right) \text{ for } i \in \{1,...,m\}$$

If $m=N-1$, then there is a solution $\lambda^{(N-1)}$ to $g^{(N-1)}(\lambda)=E$, and from Proposition 1, $\kappa^{(N-1)}$, with components given by

$$\kappa_i^{(N-1)} = f^{-1}\left(\dfrac{\lambda^{(N-1)}}{p_i}\right) \text{ for } i \in \{1,...,N-1\}$$

is the global optimum of $U^{(N-1)}$ in $V^{(N-1)}$. From Eq. (9), it follows that the components of the global optimum $\kappa^{(opt)}$ of $U$ in $S$ are given by Eq. (18).

If $m < N-1$, then $g^{(k)}(\lambda)=E$ does not have a solution for $k \in \{m+1,...,N-1\}$. So from Proposition 1, the global optimum $\kappa^{(m+1)}$ of $U^{(m+1)}$ in $V^{(m+1)}$ can be written as

$$\boldsymbol{\kappa}^{(m+1)} = \left(\kappa_1^{(m)},...,\kappa_m^{(m)}, E - \kappa_1^{(m)} - ... - \kappa_m^{(m)}\right)$$

Noting that the components of $\kappa^{(m+1)}$ add to $E$ and iteratively applying Proposition 1 gives

$$\boldsymbol{\kappa}^{(N-1)} = \left(\kappa_1^{(m)},...,\kappa_m^{(m)}, E - \kappa_1^{(m)} - ... - \kappa_m^{(m)}, 0,...,0\right)$$

From Eq. (9) it follows that

$$\kappa^{(opt)} = \left(\kappa^{(N-1)}, 0\right)$$

Since

$$g^{(m)}\left(\lambda^{(m)}\right) = \sum_{i=1}^{m+1} f^{-1}\left(\frac{\lambda^{(m)}}{p_i}\right) = E,$$

we have $E - \kappa_1^{(m)} - \ldots - \kappa_m^{(m)} = f^{-1}\left(\frac{\lambda^{(m)}}{p_{m+1}}\right)$ and it follows that the components of $\kappa^{(opt)}$ are given by Eq. (20).

(ii) Finally, if $g^{(k)}(\lambda)=E$ does not have a solution for any $k \in \{1,\ldots,N-1\}$, then we see by iteratively applying Proposition 1 as before that the global optimum, $\kappa^{(N-1)}$ of $U^{(N-1)}$ in $V^{(N-1)}$ can be written as

$$\kappa^{(N-1)} = \left(\kappa_1^{(1)}, 0, \ldots, 0\right)$$

where $\{\kappa_1^{(1)}\}$ is the global optimum of $U^{(1)}$ in $V^{(1)}$. Since $g^{(1)}(\lambda)=E$ does not have a solution, it follows from Proposition 1 that $\kappa_1^{(1)}$ must belong to $S^{(1)} = \{E\}$. Substituting in the above expression for $\kappa^{(N-1)}$ and using Eq. (9) yields Eq. (21). □

Theorem 3 specifies situations in which it is optimal to completely "drop" items: Eq. (20) describes a solution in which the $m+1$ items with the highest probing probabilities receive resource, and the remaining items receive no resource at all. Eq. (21) describes a solution in which only the item most likely to be proved receives resource.

**Existence of solutions**

We will now establish conditions under which solutions for $g^{(k)}(\lambda)=E$ exist for different values of $k$. This will lead to an iterative algorithm to find the optimum of $U$.

**Lemma 4:** We further assume that the local utility function is continuously differentiable, $u \in C^1$. Then $g^{(k)}(\lambda)=E$ has a solution if and only if there is a solution $\lambda^{(k-1)}$ to $g^{(k-1)}(\lambda) = E$ such that $\lambda^{(k-1)} < p_k f(0)$.

**Proof of Lemma 4:** Since $u$ is increasing and convex, $u' = f$ is strictly decreasing in $[0, E]$. Under the assumption $u \in C^1$, it follows $f^{-1}$ is continuous and strictly decreasing. So $g^{(k)}(\lambda)$ is also continuous and strictly decreasing for any $k$. Evaluating $g^{(k)}$ at $p_1 f(E)$ gives the following inequality

$$g^{(k)}\left(p_1 f(E)\right) = \sum_{i=1}^{k+1} f^{-1}\left(\frac{p_1 f(E)}{p_i}\right) = E + \sum_{i=2}^{k+1} f^{-1}\left(\frac{p_1 f(E)}{p_i}\right) > E \qquad (22)$$

Suppose $g^{(k)}(\lambda^{(k)})=E$ holds for some $\lambda^{(k)}$. Since

$$g^{(k)}(\lambda) = \sum_{i=1}^{k+1} f^{-1}\left(\frac{\lambda}{p_i}\right) = g^{(k-1)}(\lambda) + f^{-1}\left(\frac{\lambda}{p_{k+1}}\right) > g^{(k-1)}(\lambda), \qquad (23)$$

it follows that $E > g^{(k-1)}\left(\lambda^{(k)}\right)$. Eq. (22) holds for any $k$, so $g^{(k-1)}\left(p_1 f(E)\right) > E$. Since $g^{(k-1)}$ is continuous and strictly decreasing from the intermediate value theorem it follows $g^{(k-1)}\left(\lambda^{(k-1)}\right) = E$ for some $\lambda^{(k-1)} \in \left(p_1 f(E), \lambda^{(k)}\right)$. Furthermore, since $f$ is decreasing and $E - \kappa_1 - \ldots - \kappa_k \geq 0$, we have $p_{k+1} f\left(E - \kappa_1 - \ldots - \kappa_k\right) = \lambda^{(k)} \leq p_{k+1} f(0)$. So $p_{k+1} f(0) \geq \lambda^{(k)} > \lambda^{(k-1)}$.

We have shown that

If $g^{(k)}(\lambda) = E$ has a solution $\lambda^{(k)}$, then there is a solution $\lambda^{(k-1)}$ to $g^{(k-1)}(\lambda) = E$ such that $\lambda^{(k-1)} < p_{k+1} f(0)$ $\qquad (24)$

We will now show that the converse holds as well. Suppose there is a solution $\lambda^{(k-1)}$ to $g^{(k-1)}\left(\lambda^{(k-1)}\right) = E$ such that $p_{k+1} f(0) > \lambda^{(k-1)}$. Then

$$g^{(k)}\left(p_{k+1} f(0)\right) = \sum_{i=1}^{k+1} f^{-1}\left(\frac{p_{k+1} f(0)}{p_i}\right)$$

$$< 0 + \sum_{i=1}^{k} f^{-1}\left(\frac{\lambda^{(k-1)}}{p_i}\right)$$

$$g^{(k)}\left(p_{k+1} f(0)\right) < E$$

So, from the intermediate-value theorem, there is a solution $\lambda^{(k)} \in \left(p_1 f(E), p_{k+1} f(0)\right)$ such that $g^{(k)}\left(\lambda^{(k)}\right) = E$. $\square$

Combining Lemma 4 with Theorem 3 gives the algorithm in Theorem 1 of the main text. $\square$

**SOLVED EXAMPLES**

We now illustrate the properties of the optimal allocation strategy for two different local utility functions – one with finite $u'(0)$ (Figure 3) and the other with infinite $u'(0)$ (Figure 5).

*Example 1*. In Figure 3, we used the local utility function depicted in Figure 1, which is $u(\kappa)=1-e^{-0.1\kappa}$. Fig. 3A-B show the optimal allocation for set sizes 2 and 3 and three different values of the resource budget *E*. In all cases, optimal allocation deviates not only from an equal allocation ($\kappa_i=E/N$) but also from proportional allocation ($\kappa_i=p_iE$). Moreover, the derivative of *u* at 0 is 0.1, and as a result, it is optimal to "drop" items when the resource budget is sufficiently low (Fig. 3A-B).

To examine the benefits of using an optimal allocation strategy, we compute overall utility of all three policies in a hypothetical experiment in which one item is cued with validity $p_1$ and the remaining probing probability $1-p_1$ is divided equally among the remaining items. We find that for each of the three policies, overall utility decreases with increasing set size and increases with increasing resource budget (Fig. 3C). Both the optimal and the proportional allocation strategies produce substantially higher utility than equal allocation, of course mostly so when the probing probabilities deviate more from equality. In the optimal allocation strategy, given the same resource budget, the lowest $p_1$ for which the dropping happen decreases with increasing set size (Fig. 3D).

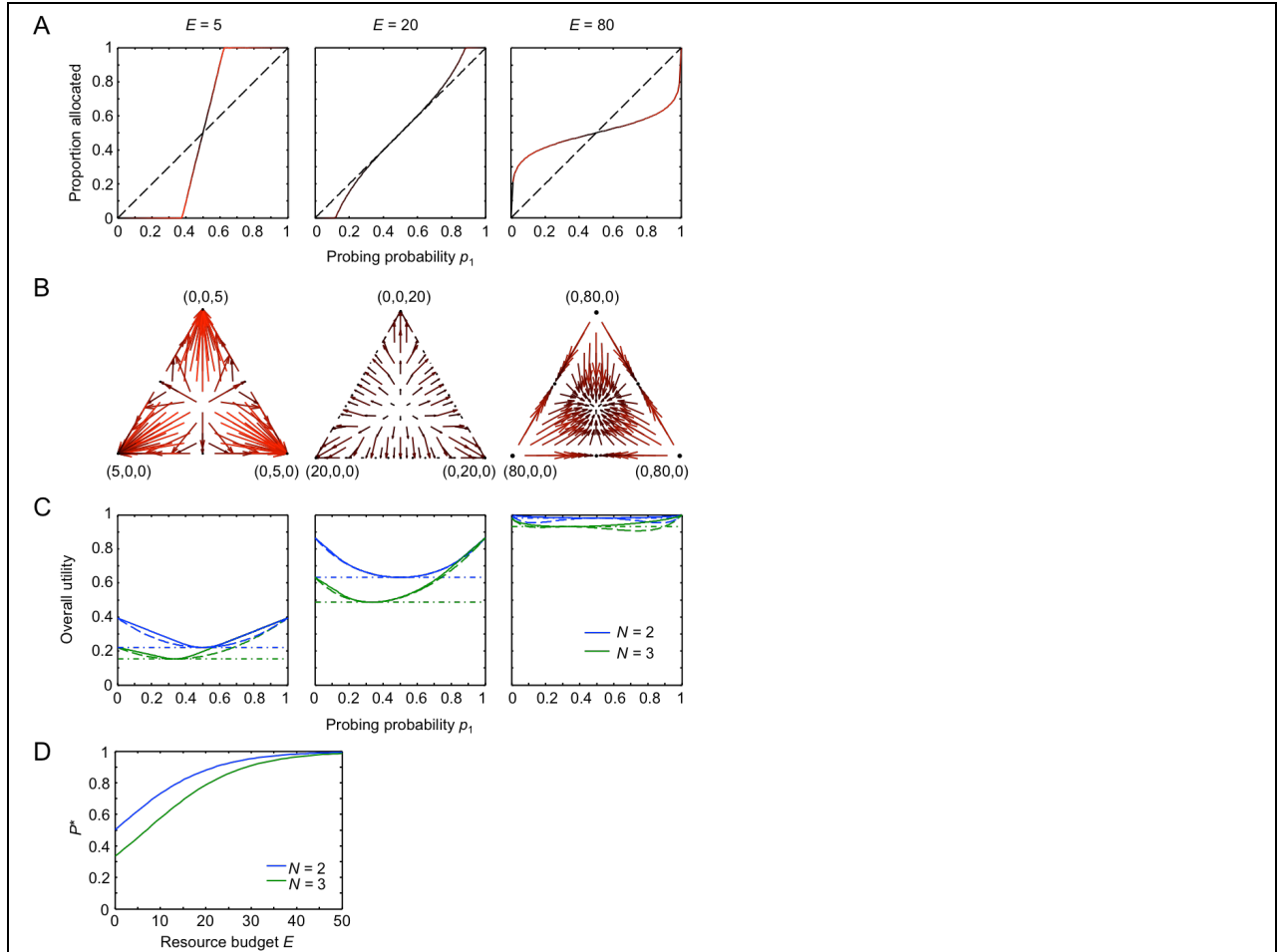

**Figure 3:** Example 1. Illustration of how the optimal allocation strategy differs from the proportional strategy for the local utility function $u(\kappa)=1-e^{0.1\kappa}$ and for different resource budgets ($E$ = 5, 20, 100). **(A)** $N$=2. The proportion of resource allocated to an item as a function of the probing probability of that item. Solid line: optimal allocation. Dashed line: proportional allocation. Item dropping corresponds to a $y$-value of 0 or 1. The red saturation of a point on the curve represents the magnitude of the deviation from proportional allocation. **(B)** $N$=3. Vectors from the point corresponding to the probing probability vector ($p_1$, $p_2$, $p_3$) to the normalized optimal resource allocation vector ($\kappa_1$, $\kappa_2$, $\kappa_3$)/$E$ on the three-dimensional simplex given by $x_1+x_2+x_3=1$, $x_1>0$, $x_2>0$, $x_3>0$. Item dropping corrresponds to an arrow that starts in the interior and ends on the border. The red saturation of an arrow represents the magnitude of the deviation from proportional allocation. **(C)** Overall utility as a function of the probing probability of the first location, assuming that all other locations have equal probability of being probed. Solid: optimal allocation. Dashed: proportional allocation. Dot-dashed: equal allocation. **(D)** The lowest probing probability ($P^*$) of the cued item for which dropping of items happens under the optimal allocation strategy, as a function of the resource budget. The probing probabilities of non-cued items are equal.

*Example 2*. Consider a task in which the observer discriminates between two stimuli, *s*=0 and *s*=1 (this could also be a detection task). For the utility function, we have to specify a noise model; we make the standard assumption of Gaussian noise, and denote its inverse variance at the $i^{th}$ location by $\kappa_i$ (*i*=1,2). Then, signal detection theory tells us that the local utility function at the $i^{th}$ location is

$$u(\kappa_i) = \frac{1}{2} + \frac{1}{2}\mathrm{erf}\sqrt{\frac{\kappa_i}{2}}. \tag{25}$$

Now, the derivative of *u* at 0 is infinite. We follow the algorithm above under the constraint that $\kappa_1+\kappa_2$ is constant. Then, the iterative condition $\lambda^{(k-1)} < p_{k+1} f(0)$ will hold for all *k*, the algorithm will update *k* until it reaches *N*, and thus, every item will receive resource: no item dropping will occur (Fig. 4A-B). As in Example 1, overall utility under each strategy decreases with increasing set size and increases with increasing resource budget (Fig. 4C). For the parameter values tested, the optimal and proportional allocation policies are practically indistinguishable, but both are substantially better than equal allocation (Fig. 4C).

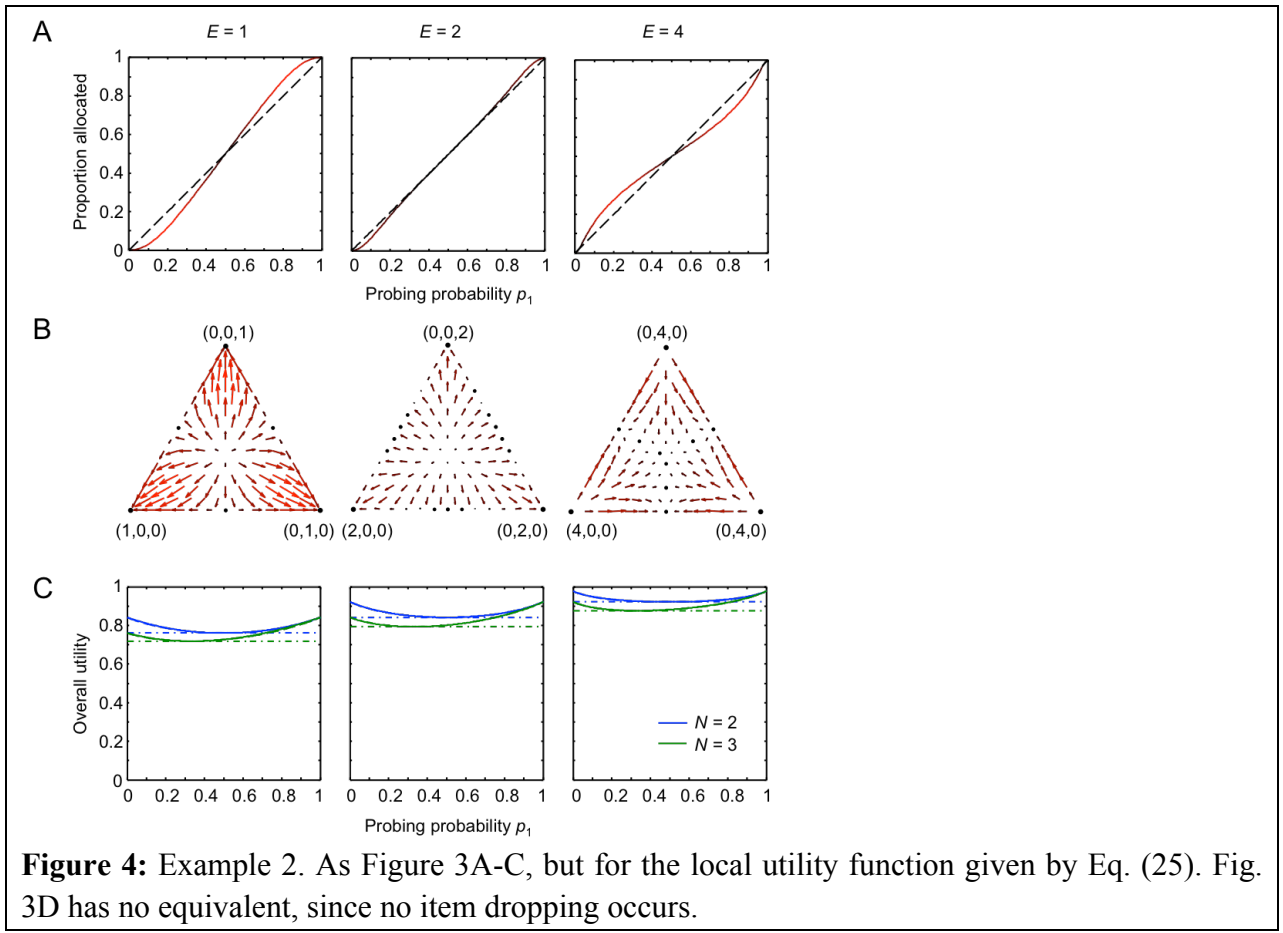

**Figure 4:** Example 2. As Figure 3A-C, but for the local utility function given by Eq. (25). Fig. 3D has no equivalent, since no item dropping occurs.

**DISCUSSION**

Although attentional cueing is one of the most common manipulations in cognitive psychology, the way to optimally allocate attention in response to an attentional cue is unknown. One specific case has been studied (Bays, 2014), but a general theory is lacking. Here, we have formulated the optimal strategy for allocating a fixed budget of attentional resource among multiple stimuli in tasks that satisfy three intuitive properties (separability, equivalence, and diminishing returns). We proved that there exists a unique optimal allocation, derived an efficient algorithm to find this optimum, and compared the optimal strategy with alternative ones. The efficiency of the algorithm consists of reducing an ($N$-1)-dimensional optimization problem to a series of at most $N$ one-dimensional problems. It is possible that this efficiency could help the brain attain a solution that is close to optimal.

We found that allocating attention in proportion to the probing probabilities is not optimal, although in many cases close to optimal. In particular, when the derivative of the local utility function at zero is finite, as it is in many cued estimation tasks, and the amount of resource is low enough, it is optimal to allocate zero resource to items that have a nonzero probability of being probed. Of course, the brain might not be able to allocate strictly zero resource to an item; this remains to be determined.

The theory has the potential for extension. First, one could consider tasks for which the overall utility function is not separable, such as ones in which the stimuli are correlated. A particularly important category of such tasks are ones in which the subject makes a "global" judgment, involving the whole set of stimuli, such as detecting a single target among multiple items. There have been several studies that used attentional cues in such tasks (Giordano, McElree, & Carrasco, 2009; Shimozaki, Schoonveld, & Eckstein, 2012). Second, further work could consider relaxation of the independence assumption. Third, it would be interesting to replace the fixed resource budget $E$ by a "soft constraint", where $U(\kappa)$ is replaced by $U(\kappa)-cE$, where $c$ is a constant (Van den Berg & Ma). Both terms depend on $\kappa$, and the cost of investing more resource (increasing $E$) could be offset by performance gains (an increase in $U(\kappa)$).

The present work might serve as a first step in a more rigorous characterization of the allocation of attention. If, in some tasks, it is found that subjects allocate attention optimally, it will be important to determine a plausible neurobiological mechanism for such a strategy. A good start could be the approach by Paul Bays (Bays, 2014): a neural population model in which the probing probabilities influence the amplitudes of the population activities encoding the items.